\magnification=1200
\hsize =34true pc\vsize = 48true pc
\hoffset=.375 true in                                                                                             
\voffset=.5true in
\font\mybig=cmbx12 at 12pt

\font\mysmall=cmr8 at 8 pt
\font\eightit=cmti8

\def\c{\bf C}

\def\f{\bf F}

\def\q{\bf Q}

\def\z{\bf Z}

\def\r{\bf R}

\def\hh{\cal H}

\def\ff{\cal F}

\def\ss{\cal S}

\def\bb{\cal B}
\def\picture #1 by #2 (#3){
		\vbox to #2{
		\hrule width #1 height 0pt depth 0pt
		\vfill
		\special {picture #3}}}
\font\teneufm eufm10 
\font\seveneufm eufm7 
\font\fiveeufm eufm5
\newfam\eufm
\textfont\eufm\teneufm
\scriptfont\eufm\seveneufm
\scriptscriptfont\eufm\fiveeufm
\def\frak#1{{\fam\eufm\teneufm#1}}
\font\mybig=cmbx12 at 12pt 
\def\picture #1 by #2 (#3){
		\vbox to #2{
		\hrule width #1 height 0pt depth 0pt
		\vfill
		\special {picture #3}}}
\centerline {\bf \mybig Airy functions over local fields\/}
\vskip 0.5 true in
		\centerline {\bf Rahul N. Fernandez, V. S. Varadarajan and David Weisbart}
\vskip 0.5 true in\noindent 
\hfill {\eightit To the memory of Moshe Flato}
\vskip 0.5 true in
{\mysmall
\centerline {ABSTRACT}
\vskip 0.5 true in
Airy integrals are very classical but in recent years they have been generalized to higher dimensions and these generalizations have proved to be very useful in studying the topology of the moduli spaces of curves. We study a natural generalization of these integrals when the ground field is a non-archimedean local field such as the field of p-adic numbers. We prove that the p-adic Airy integrals are locally constant functions of moderate growth and present evidence that the Airy integrals associated to compact p-adic Lie groups also have these properties.}
\vskip 0.5 true in\noindent
{\bf 0. Introduction.\/} It is now ten years since Moshe Flato passed away in a sudden and untimely fashion. He had an extraordinarily broad point of view in theoretical physics and pioneered many ideas which became fashionable decades later--deformation quantization, physics in conformal space, singletons, and so on. This paper is a small contribution dedicated to his memory. 
\medskip
In [K] Kontsevich proved certain conjectures of Witten [Wi] about  intersection theory in the moduli spaces of curves, which arose from two-dimensional gravity. For this purpose the notion of matrix Airy integrals was  introduced in [K]; see [L] where even more general integrals are considered. The Kontsevitch integrals were generalized considerably in [FV] where the unitary group ${\rm U}(N)$ and the space of hermitian matrices ${\hh}(N)$ occurring in [K] were replaced by a compact connected real Lie group $G$ and its Lie algebra $\frak g$. It now appears that these general matrix Airy integrals may have an arithmetic side to them. In this paper we explore this idea and study the Airy integrals over a {\it local non-archimedean field\/}. We hope that $p$-adic Airy integrals may also have connections with moduli spaces. They are a part of  non-archimedean physics which has been of interest since the appearance of the path-breaking papers of Volovich [V1] [V2] introducing the hypothesis that the geometry of space-time in sub-Planckian regimes is non-archimedean. For some consequences of this hypothesis for particle classification see [Va] [Vi].
\bigskip\noindent  
{\bf 1. Airy functions.\/} Let $X$ be a t.d.~space, i.e., a Hausdorff space with the property that the compact open sets form a base for the topology. We write ${\ss}{\bb}(X)$ for the space of {\it Schwartz-Bruhat functions\/}, i.e., locally constant complex functions on $X$ with compact support. {\it Any\/} linear functional ${\ss}{\bb}(X)\longrightarrow {\c}$ is called a {\it distribution\/} on $X$ (see [H]). Let $\mu$ be a Borel measure on $X$ and $F$ a locally $\mu$-integrable Borel function. Then $F$ defines the distribution
$$
T_F: f\longrightarrow \int_XFf\,{\rm d}\mu\qquad (f\in {\ss}{\bb}(X))
$$
and $T_F$ determines $F$ $\mu$-almost everywhere. If $T$ is a distribution, we say that $T$ is a locally integrable function with respect to $\mu$ if  $T=T_F$ for some $F$.
\medskip 
Let $V$ be a finite-dimensional vector space over a local non-archimedean field $K$. We denote by ${\rm d}V$ (or ${\rm d}x, \,{\rm d}y$ etc) any Haar measure on $V$.  We say that a distribution on $V$ is a locally integrable function if it is locally integrable with respect to Haar measure. We also choose a non-trivial additive character $\psi$ on $K$; any other character of $K$ is of the form $y\longmapsto \psi (cy)$ for some $c\in K^\times$. On ${\q}_p$, we use the $p$-adic expansion $\sum _{n>>-\infty}a_np^n$ for any element of ${\q}_p$, where $a_n\in \{0,1,2,\dots ,p-1\}$ for all $n$, and write any $x\in {\q}_p$ as $x=r+y$ where $y\in {\z}_p$, the ring of $p$-adic integers, and $r\in {\z}[p^{-1}]$. Then $r$ is determined up to addition by an integer and so $e^{2\pi ir}$ is uniquely defined. We set $\psi_0(x)=e^{2\pi ir}$ to obtain a non-trivial additive character of ${\q}_p$. If $K$ is a finite extension of ${\q}_p$ for some $p$, we define $\psi(x)=\psi_0({\rm Tr}(x))$, where ${\rm Tr}={\rm Tr}_{K/{\q}_p}$. In what follows, the choice of $\psi$ is immaterial as long as it is non-trivial. If $K$ has characteristic $p>0$, we choose the additive character $\psi$ as follows. We know that $K\simeq K_f$, the field of Laurent series $s$ over the finite field ${\f}_q (q=p^f)$ of $q$ elements,
$$
s=\sum _{r>>-\infty}s(r)T^r\qquad (T  \hbox { an indeterminate}, s(r)\in {\f}_q).
$$ 
Let $\theta$ be a non-trivial additive character of ${\f}_p$ and let ${\rm Tr}$ denote the trace from ${\f}_q$ to ${\f}_p$. Since there are elements $a\in {\f}_q$ with ${\rm Tr}(a)\not=0$, we see that $\theta_q := \theta\circ {\rm Tr}$ is a non-trivial additive character of ${\f}_q$. Define the linear map ${\rm Res}$ from $K_f$ to ${\f}_q$ by ${\rm Res}(s)=s(-1),\, s\in K_f$. 
Then $
\psi : s\longmapsto \theta_q({\rm Res}(s))=\theta ({\rm Tr}({\rm Res}(s)))$ is a non-trivial additive character on $K_f$. 
\medskip
The field $K$ has a canonical valuation $|{\cdot}|$ defined by ${\rm d}(ax)=|a|{\rm d}x$ where ${\rm d}x$ is a Haar measure on $K$. Let $R$ be the compact open ring of integers of $K$, $P$ its maximal ideal, $\varpi$ a uniformisant, i.e., $P=R\varpi$, and $k=R/P$, the residue field of $K$. We have $|\varpi|=q^{-1}$ where $q=|k|$. As $\psi$ is trivial on $P^{-r}=R\varpi^{-r}=\{|x|\le q^{r}\}$ for some $r<<0$, we can speak of the largest integer $r$ such that $\psi$ is $1$ on $P^{-r}$; this integer is the order of $\psi$ and is denoted by ${\rm ord}(\psi)$. If $c\not=0$ is an element of $K$ with $|c|=q^s$, $\xi$ is a non-trivial additive character of $K$, and $\eta$ is  defined by $\eta (x)=\xi(cx)$, then ${\rm ord } (\eta)={\rm ord } (\xi)-s$. Finally we write
$$
U=\{u\in K||u|=1\}.
$$
For all of this see [W].
\medskip
Fourier transforms of objects on $V$ are similar objects on the dual $V^\prime$ of $V$. However it is more practical for us (although less canonical) to choose a non-singular bilinear form $V\times V\longrightarrow K$, denoted by $(x, y)$, and define, for any $f\in {\ss}{\bb}(V)$, its Fourier transform $\widehat f$ by 
$$
\widehat f(x)=\int _Vf(y)\psi(-(x, y))\,{\rm d}y\qquad (x\in V).
$$
It is  well-known that $\widehat f\in {\ss}{\bb}(V)$ and that for a suitable normalization of ${\rm d}V$ (self-dual Haar measure) we have, for all $f\in {\ss}{\bb}(V)$,
$$
f(y)=\int_V\widehat f(x)\psi((x, y))\,{\rm d}x\qquad (y\in V).
$$
Thus the {\it Fourier transform map\/} ${\ff} : f\longmapsto \widehat f$  
is a linear isomorphism of ${\ss}{\bb}(V)$ with itself that takes multiplication into convolution and vice versa, and satisfies ${\ff}^2f(x)=f(-x),\  {\ff}^4f=f$. Once ${\ff}$ is defined, its definition can be extended to distributions by duality:
$$
\widehat T(f)=T(\widehat f)\qquad (f\in {\ss}{\bb}(V)),\qquad {\ff}T=\widehat T.
$$
In particular, if $T=t$ d$x$, then $\widehat T=\widehat t$ d$x$.
\medskip
Let $h$ be a polynomial function on $V$ with coefficients in $K$. Then the function $\psi \circ h : y\longmapsto \psi(h(y))$ is bounded and locally constant and so defines a distribution on $V$. Its Fourier transform is called the {\it Airy distribution\/} defined by $h$ and is denoted by $A_h$. We say that $h$ has the {\it Airy property\/} if $A_h$ is a locally integrable function with at most polynomial growth, i.e., $A_h(x)=O(|x|^s)$ for some $s\ge 0$ as $|x|\to \infty$.  Here we imitate the definition when the ground field is ${\r}$. If $h$ has the Airy property, we write $A_h(x)$ for the corresponding locally constant function and call such functions {\it Airy functions\/}. 
\medskip
In order to study the Airy distributions, it is convenient to take an alternative approach and define the Airy function as an {\it improper Riemann integral\/} over $V$. If $f$ is a continuous function on $V$, we define the {\it improper Riemann integral\/} of $f$ over $V$ relative to the sequence $(K_n)$ of compact open subsets of $V$ with $K_n\subset K_{n+1}, \cup _nK_n=V$ as 
$$
R\int_{V}f(y)\,{\rm d}y:=\lim_{n\to\infty}\int_{K_n}f(y)\,{\rm d}y
$$
if this limit exists. Let $|{\cdot}|$ be a non-archimedean norm on $V$. It follows from [W] (Proposition 3, p. 26) that the values of $|{\cdot}|$ on $V\setminus \{0\}$ form a discrete subset of ${\r}\setminus \{0\}$ and so the norm values $>1$ can be written as a sequence $b_1<b_2<\dots, b_r\to \infty$ as $r\to \infty$. We define
$$
R\int_{V}f(y)\,{\rm d}y:=\lim_{r\to\infty}\int_{|y|\le b_r}f(y)\,{\rm d}y
$$
if the limit exists. It will exist if and only if the series $\sum _{r=1}^\infty \int _{|y|=b_r}f(y)\,{\rm d}y$ is convergent, and then
$$
R\int_{V}f(y)\,{\rm d}y=\int_{|y|\le 1}f(y)\,{\rm d}y+\sum_{1\le r<\infty}\int_{|y|=b_r}f(y)\,{\rm d}y.
$$
We will use this concept to define the Airy function directly. This is analogous to defining the Airy integral in the real case as the improper Riemann integral which is the principal value of
$$
\int_{-\infty}^{+\infty}\cos(y^3-xy) \,{\rm d}y,
$$ 
just as Airy himself did [A] [FV]. We shall in fact show that under suitable conditions on the polynomial $h$ on $V$,
$$
A(x)=\int _{|y|\le 1}\psi \big (h(y)-(x,y)\big )\,{\rm d}y+\sum _{r=1}^\infty \int _{|y|=b_r}\psi\big (h(y)-(x,y)\big )\,{\rm d}y
$$
is well-defined for all $x\in V$, is locally constant with polynomial growth (at most) at infinity, and that the distribution defined by $A$ is the Airy distribution $A_h$. 
\medskip
If $E$ is any compact and open subset of $V$, the integral
$$
A_E(x):=\int_E\psi\big (h(y)-(x,y)\big )\,{\rm d}y
$$
is the Fourier transform of the Schwartz-Bruhat function $1_E(y)\psi(h(y))$, where $1_E$ is the characteristic function of $E$, and so is a Schwartz-Bruhat function. So the first term and the remaining terms occurring in the series on the right side of the definition of $A(x)$ are individually all Schwartz-Bruhat functions. The key now is to prove that for any fixed integer $B>0$, there exists $r_0=r_0(B)\ge 0$ such that 
$$
\int_{|y|=b_r}\psi\big (h(y)-(x,y)\big )\,{\rm d}y=0\qquad (r\ge r_0, |x|\le B)\eqno (\ast\ast).
$$
Suppose that this has been established. Then $A$ is well-defined and locally constant. Then for any $r_1\ge r_0$ and any $f\in {\ss}{\bb}(V)$ with support of $f$ contained in $\{x\ |\ |x|\le B\}$, 
$$
\eqalign {\int _{V}A(x)f(x)\,{\rm d}x&=\int _{V}f(x)\bigg (\int _{|y|\le b_{r_1}}\psi\big (h(y)-(x,y)\big )\,{\rm d}y\bigg )\,{\rm d}x\cr
&=\int_{|y|\le b_{r_1}}\psi(h(y))\bigg (\int _{V}f(x)\psi(-(x,y))\,{\rm d}x\bigg )\,{\rm d}y\cr
&=\int_{|y|\le b_{r_1}}\psi(h(y))\widehat f(y)\,{\rm d}y\cr
&=\int _{|y|\le 1}\psi(h(y))\widehat f(y)\,{\rm d}y+\sum_{1<r\le r_1} \int _{|y|=b_r}\psi(h(y))\widehat f(y)\,{\rm d}y.\cr}
$$
But $\widehat f$ itself has compact support and so we can choose $r_1\ge r_0$ such that $\widehat f(y)$ is $0$ if $|y|>b_{r_1}$. Hence
$$
\int _{V}A(x)f(x)\,{\rm d}x=\int_{|y|\le b_{r_1}}\psi(h(y))\widehat f(y)\,{\rm d}y=\int_{V}\psi (h(y))\widehat f(y)\,{\rm d}y
$$
showing that $A$ is in fact the Airy distribution defined by $h$. So one has to prove $(\ast\ast)$. Actually, in our applications we shall prove $(\ast\ast)$ in a much stronger form, which will lead to a bound at infinity for $A(x)$. 
Indeed we shall prove $(\ast \ast)$ for $|x|\le B(r)$ where $B(v)$ is an increasing function going to $\infty$. Also write $b_r=b(r)$ where $b$ is an increasing function going to $\infty$. For bounding $A(x)$ we may thus assume that $|x|>B(r_0)$. Let $r>r_0$ be the smallest integer such that $|x|\le B(r)$. Then $|x|>B(r-1)$. Let $\beta$ be the function inverse to $B$. Then $r-1<\beta(|x|)$. But $r-1\ge r_0$, and so we have 
$$
|A(x)|\le \int _{|y|\le b(r-1)}\,{\rm d}y\le Cb(r-1)^m\le Cb(\beta(|x|))^m.
$$
If for some constants $L, M, c, a>0$ we have $b(r)=Lc^r$ and $B(r)=Mc^{ar}$, this leads to the estimate $|A(x)|=O\big (|x|^{m/a}\big)$.
\bigskip\noindent
{\bf 2. The statement of the main theorems.\/} We have the following theorem. For the quadratic case in dimension 1 see [VVZ]. We work on $V=K^m$ with the norm $|v|=\max_i|v_i|$ for $v=(v_1, \dots ,v_m)$.
\bigskip\noindent
{\bf Theorem 1.\/} {\it  Let $h$ be a polynomial on $K^m$ of degree $n\ge 2$ and $h_n$ its homogeneous part of degree $n$. Assume that $h_n$ has the following form
$$
h_n(y)=a_1y_1^n+a_2y_2^n+\dots +a_my_m^n\qquad (a_i\not=0\ \forall i).
$$
If either $K$ has characteristic $0$, or has characteristic $p>0$ where $p$ does not divide $n$, then $h$ has the Airy property. The function
$$
A(x)=R\int _{K^m}\psi\big (h(y)-(x,y)\big )\,{\rm d}y
$$
is well-defined for all $x\in K$, is locally constant, and is $O(|x|^{m/(n-1)})$ for $|x|\to \infty$. Moreover, the distribution defined by $A$ is the Airy distribution $A_h$. In particular these results are true when $m=1$ for any polynomial of degree $\ge 2$. If $m=1$ and $\deg(h)\le 1$, then the Airy distribution $A_h$ is a delta function.\/}
\bigskip
Suppose now that $K=K_f$ has characteristic $p>0$ and we assume $p$ divides $n$. Here we do not have a true analogue to Theorem 1 and have to restrict ourselves to the case of one variable, $m=1$. For any $c\in K$ we write
$$
 Qc=c^\sharp =\sum_jc^\sharp (j)T^j,\qquad c^\sharp(j)=(c(-1+p(j+1)))^{p^{-1}}\qquad (c\in K_f).
$$
The map $Q : c\longmapsto c^\sharp$ is additive from $K_f$ to itself. For any polynomial $h(y)$ with coefficients in $K_f$ we write $h^\sharp$ for the polynomial obtained by the process of replacing each term $cy^{mp^r}$ in $h$ with $c\not=0, r\ge 1, (m,p)=1$, by $(Q^rc) y^m$:
$$
cy^{mp^r}\longmapsto  (Q^rc)y^m.
$$
It is clear that in $h^\sharp$ every term has degree not divisible by $p$ and  that if $h$ is of degree $n$ and $(n,p)=1$, then $h^\sharp$ has degree $n$ also and the same leading term as $h$. The result in characteristic $p$ takes the following form.
\bigskip\noindent
{\bf Theorem 2.\/} {\it Let $\psi=\theta_q\circ {\rm Res}$ be as above and $h$ be a polynomial over $K_f$. Then $\psi(h)=\psi(h^\sharp)$. Moreover, if $h^\sharp$ has degree $n^\sharp\ge 2$, then $h$ has the Airy property and the distribution $A_h$ is the one defined by 
$$
A(x)=R\int _K\psi(h^\sharp(y)-xy)\,{\rm d}y
$$
which is well-defined and locally constant on $K$ and $O\big (|x|^{1/(n^\sharp -1)}\big )$ as $|x|\to \infty$. In particular these results are true for $h$ itself if the degree of $h$ is prime to $p$. If $\deg(h^\sharp)\le 1$ then the Airy distribution defined by $h$ is a delta function.\/}
\medskip
\bigskip\noindent
{\bf Idea of the proof in characteristic $0$.\/} We look at the case of one variable. Let $h(y)=y^n+c_1y^{n-1}+\dots +c_{n-1}y$. Making the change of variable $y=\varpi^{-r}z$,
$$
\int _{|y|=q^r}\psi\big (h(y)-xy\big )\,{\rm d}y=q^r\int_{|z|=1}\psi\big (\varpi^{-nr}\left (h_r(z)\big)\right )\,{\rm d}z
$$
where
$$
h_r(z)=z^n+\varpi^r c_1 z^{n-1}+\varpi^{2r}c_2z^{n-2}+\dots +\varpi^{(n-1)r} (c_{n-1}-x)z.
$$
When $r\to \infty$ all the coefficients of $h_r$ except the leading one become small and so for large $r$ we may regard $h_r$ as a {\it small perturbation\/} of the function $z^n$. The key step is therefore a study of the polynomial  
$$
F(z : u)=z^n+u_1z^{n-1}+u_2z^{n-2}+\dots +u_{n-1}z\quad (u=(u_1, u_2, \dots , u_{n-1})\in R^{n-1})
$$
on $U=\{|z|=1\}$ {\it when the parameters $|u_i|$ are small\/}. This will allow us to simplify the integral in question. We write $U^r=\{t^r\ |\ t\in U\}$ 
and $z\longmapsto \bar z$ for the map $R\longrightarrow R/P=k$. When the number of variables is greater than one, we evaluate the integrals one variable at a time and so it is necessary to study the one variable case where the coefficients of $h$ are not fixed but vary.
\medskip
\bigskip\noindent
{\bf 3. Structure of $F(z : u)$ when $|u|$ is small.\/} We begin with a proposition.
\bigskip\noindent
{\bf Proposition 1.\/} {\it Let $K$ have arbitrary characteristic. Suppose that $p$ does not divide $n$. Let $|u_i|\le q^{-1}$ for all $i$. Then the equation $F(z : u)=t^n$ has, for each $z\in U$, a unique solution $t=t(z)\in U$ with $\bar t=\bar z$, and the map $z\longmapsto t(z)$ is an analytic diffeomorphism of $U$ with itself with $|{\rm d}t/{\rm d}z|=1$.\/}
\bigskip\noindent
{\bf Proof.\/} Given $z\in U$, the polynomial $T^n-F(z : u)$ in the indeterminate $T$ goes over to $T^n-{\bar z}^n$ mod $P$ and has the simple root $T=\bar z$. By Hensel's lemma we can lift $\bar z$ to a unique root $t\in R$, so that $\bar t=\bar z$ and $t^n=F(z : u)$, and clearly $|t|=1$. The map $z\longmapsto t(z)$ is thus well-defined. Suppose that $F(z_i : u)=t^n$ for $i=1,2$ with $\bar t=\bar z_1=\bar z_2$, but $z_1\not=z_2$. Then
$$
z_1^n-z_2^n+u_1(z_1^{n-1}-z_2^{n-1})+\dots + u_{n-1}(z_1-z_2)=0
$$
so that, dividing by $(z_1-z_2)$ we get, as the $u_i\in P$ for all $i$, $
(z_1^{n-1}+z_1^{n-2}z_2+\dots +z_2^{n-1})\in P$. So, writing $z_1=zz_2$ where $\bar z=1$ we get $z^{n-1}+z^{n-2}+\dots +1\in P$. Since $z\equiv 1$ mod $P$, this implies that $n\equiv 0$ mod $P$, a contradiction. To show that the map $t$ is onto, let $t\in U$ be given. By Hensel's lemma applied to the polynomial $F(T : u)-t^n$, we can find a $z\in R$ with $\bar z=\bar t$ and $F(z : u)=t^n$. It is then immediate that $t(z)=t$. 
\medskip
It only remains to show that $t$ is analytic. For, if we have shown this, we can differentiate $t$ to get, remembering that $|n|=1$, 
$$
|nt^{n-1}({\rm d}t/{\rm d}z)|=|nz^{n-1}+(n-1)u_1z^{n-2}+\dots +u_{n-1}|=1,
$$
giving $|{\rm d}t/{\rm d}z|=1$. So by the inverse function theorem, $t$ will be an analytic diffeomorphism. For the analyticity we just have to find {\it some\/} analytic map $f(U\longrightarrow U)$ such that $F(z : u)=f(z)^n$ and $f(z)\equiv z$ mod $P$ for $z\in U$. Write $f(z)=zg(z)$ so that the equation becomes $g(z)^n=1+v$ where $v=u_1z^{-1}+u_2z^{-2}+\dots +u_{n-1}z^{-(n-1)}$. If we can find an analytic map $b : P\longrightarrow 1+P $ such that $b(v)^n=1+v$, we can take $g(z)=b(u_1z^{-1}+\dots +u_{n-1}z^{-(n-1)})$. If the characteristic of $K$ is $0$ we can take $b(v)=(1+v)^{1/n}$ given by the binomial series; but this will not work if the characteristic of $K$ is $p>0$. So we construct the power series for $b$ directly. Write $b(v)=1+b_1v+b_2v^2+\dots$ where the coefficients $b_i$ are to be determined so that $b(v)^n=1+v$. If we write $b_0=1$, then the $b_r$ are uniquely determined recursively by
$$
nb_1=1,\quad nb_r=-\sum_{\sum j_\mu=r, j_\mu\le r-1}b_{j_1}b_{j_2}\dots b_{j_n} \quad (r\ge 2)
$$
with $b_0=1$. Since $p$ does not divide $n$, $b_1=n^{-1}$ is in $R$, so that $|b_1|\le 1$. It is immediate by induction on $r$ that $|b_r|\le 1$ for all $r$. Hence the power series for $b$ converges on $P$. This completes the proof.\medskip
When $p$ divides $n$, we assume that the characteristic of $K$ is $0$; $p$ is now the characteristic of the residue field $k=R/P$.
\bigskip\noindent
{\bf Lemma 2 (ch. $K=0$).\/} {\it If $p$ is odd and $n$ arbitrary, then, for any $y\in R$ with $y\not=1, y\equiv 1$ mod $p$, we have
$$
\bigg |{y^n-1\over y-1}\bigg |=|n|.
$$
This result is still true for $p=2$ and $n$ arbitrary, if $y\in R$ with $y\not=1$ but $y\equiv 1$ mod $4$.\/}
\bigskip\noindent
{\bf Proof.\/} Let $p$ be odd. We first assume that $p$ does not divide $n$. Clearly we may assume $n\ge 2$. Let $y=1+gp$ where $g\in R, g\not=0$. Then
$$
{y^n-1\over y-1}=n+{n \choose 2}gp+\dots +{n \choose n-1}(gp)^{n-2}+(gp)^{n-1}
$$
and we are done as all terms on the right but the first are in $Rp$, with $n$ a unit. 
\medskip
Let $n=p^r$ where $r\ge 1$. We use induction on $r$. Let $r\ge 2$ and assume the result is true for $1,2,\dots , r-1$. Then, writing $y_r=y^{p^r}$ we have $y_{r-1}\not=1$ and
$$
\bigg |{y_r-1\over y-1}\bigg |=\bigg |{y_{r-1}^p-1\over y_{r-1}-1}\bigg |\ \bigg |{y_{r-1}-1\over y-1}\bigg |=|p^{r-1}||p|=|p^r|.
$$
So it remains to treat the case $r=1$. Write $y=1+gp$ where $g\in R, g\not=0$. Then
$$
{y^p-1\over y-1}=p+{p \choose 2}gp+\dots +{p \choose p-1}(gp)^{p-2}+(gp)^{p-1}.
$$
All the terms except the first are in $Rp^2$ (since $p\ge 3$), so that the norm of the right side is $|p|$.
\medskip
For arbitrary $n=p^rm$ with $(m, p)=1$ we may assume that $r\ge 1, m\ge 2$. Then $y_r\equiv 1 (p)$ and $y_r\not=1$ from the preceding, and
$$
\bigg |{y^n-1\over y-1}\bigg |=\bigg |{y_r^m-1\over y_r-1}\bigg |\ \bigg |{y_r-1\over y-1}\bigg |=|m||p^r|=|n|.
$$
\medskip
We now suppose that $p=2$. The treatment is exactly the same as before. Write $y_r=y^{2^r}$. Then, for $y\not=1, y\equiv 1$ mod $4$, we have $y=1+4g$ where $g\in R, g\not=0$ and
$$
\bigg |{y^2-1\over y-1}\bigg |=|y+1|=|2+4g|=|2|.
$$
By induction we get the result as before for arbitrary $r$. The argument for $n$ odd and for $n=m2^r$ with $m$ odd, are the same. This completes the proof of the lemma.
\medskip
Define the compact open subgroups $V_p$ of $U$ by 
$$
V_p=\cases {1+Rp & if $p$ is odd\cr 1+Rp^2=1+4R & if $p=2$\cr}
$$
and let $V_p^n=\{t^n|t\in V_p\}$. 
\bigskip\noindent
{\bf Proposition 3 (ch. $K=0$).\/} {\it We have the following.
\medskip\itemitem {$(a)$} If $|u_i|<|n|$ for all $i$, then $F_u : z\longmapsto F(z : u)$ is one-one on $V_p$ and $|{\rm d}F_u/{\rm d}z|=|n|$ for $z\in V_p$.
\smallskip\itemitem {$(b)$} There exists an integer $a\ge 1$ such that if $|u_i|\le q^{-a}$ for all $i$, then $F_u$ is an analytic diffeomorphism of $V_p$ onto $V_p^n$ with $|{\rm d}F_u/{\rm d}z|=|n|$.
\smallskip}
\bigskip\noindent
{\bf Proof.\/} (a) Let $z_i\in V_p$ be such that $F(z_1 : u)=F(z_2 :u)$ with $z_1\not=z_2$. Then
$$
{z_1^n-z_2^n\over z_1-z_2}+u_1{z_1^{n-1}-z_2^{n-1}\over z_1-z_2}+\dots +u_{n-1}=0.
$$
Since $\bar z_1=\bar z_2=1$ we can write $z_1=zz_2$ where $z\not=1, \bar z=1, z\equiv 1(V_p)$ and the above equation becomes
$$
z_2^{n-1}{z^{n-1}-1\over z-1}+u_1z_2^{n-2}{z^{n-1}-1\over z-1}+\dots +u_{n-1}=0.
$$
By Lemma 2 the first term has norm $|n|$ while the others have norm $<|n|$. Hence the norm of the above expression is $|n|$ which is a contradiction. Moreover 
$$
{{\rm d}F_u\over {\rm d}z}=nz^{n-1}+g, \qquad g=u_1(n-1)z^{n-2}+\dots +u_{n-1}.
$$
Since $|g|<|n|$ and $|nz^{n-1}|=|n|$ we have $|{\rm d}F_u/{\rm d}z|=|n|$.
\medskip\noindent
(b) It is a question of proving that for a suitable integer $a\ge 1$, $F_u$ maps $V_p$ onto $V_p^n$ provided $|u_i|\le q^{-a}$ for all $i$. In view of (a), if we increase $a$ so that $q^a>|n|$, $F_u$ will also be one-one on $V_p$ and will have non-zero differential everywhere on $V_p$. It will then be an analytic diffeomorphism of $V_p$ with $V_p^n$.
\medskip
Let $a$ be arbitrary for the moment and all $|u_i|\le q^{-a}$. We first want to choose $a$ so that $F_u(V_p)\subset V_p^n$. Given $z\in V_p$ we wish to find a $t\in V_p$ such that $F_u(z)=t^n$. Write $t=z\tau$ so that the required $\tau$ should satisfy
$$
\tau\equiv 1({\rm mod }\ V_p),\qquad 1+u_1z^{-1}+\dots +u_{n-1}z^{-(n-1)}=\tau^n.
$$ 
Writing $v=u_1z^{-1}+\dots +u_{n-1}z^{-(n-1)}$ so that $|v|\le q^{-a}$ if $|u_i|\le q^{-a}$ for all $i$, we wish to solve $\tau \equiv 1 (V_p),\,\tau^n=1+v$ if $|v|\le q^{-a}$ for a suitable integer $a\ge 1$. We wish to show that the binomial series for $\tau=(1+v)^{1/n}$ converges for $|v|\le q^{-a}$, so providing the solution.
\medskip
So
$$
\tau=1+\sum _{s\ge 1}{1/n \choose s}v^s.
$$
Let $d$ be the degree of $K$ over ${\q}_p$. Then, with $n=p^rm$ where $p$ does not divide $m$, 
$$
\bigg |{1/n \choose s}\bigg |\le p^{rsd}|s!|^{-1}\le p^{sdr+sd/(p-1)}\le p^{sd(r+1)}
$$
giving
$$
\bigg |{1/n \choose s}v^s\bigg |\le p^{-2sd} \qquad (|v|\le p^{-ds(r+3)}).
$$
Now $q=p^f, d=ef$ so that $p^d=q^e$ and the above condition becomes $|v|\le q^{-e(r+3)}$. Moreover, 
$$
|\tau -1|\le \sup_{s\ge 1}\bigg |{1/n \choose s}v^s\bigg |\le \sup_{s\ge 1}\ p^{-2sd}\le p^{-2d}
$$
so that $\tau \equiv 1 (V_p)$. This finishes the argument that $F_u(V_p)\subset V_P^n$.
\medskip
We now prove that this range is exactly $V_p^n$. Let $t\in V_p$; we want to find $z\in V_p$ such that $F_u(z)=t^n$; writing $z=t\zeta$ we want to find $\zeta$ such that, with $v_i=u_it^{-i}$,
$$
\zeta\equiv 1 (V_p), \qquad \zeta^n+v_1\zeta^{n-1}+\dots +v_{n-1}\zeta =1
$$
when $|v_i|\le q^{-a}$ for all $i$, $a$ being a suitable integer $\ge 1$. Let
$$
h(v, \zeta)=\zeta^n+v_1\zeta^{n-1}+\dots +v_{n-1}\zeta-1,\qquad v=(v_1, \dots ,v_{n-1}).
$$
Then
$$
{\rm d}h=(n\zeta^{n-1}+(n-1)v_1\zeta^{n-2}+\dots +v_{n-1}){\rm d}\zeta+(\zeta^{n-1}{\rm d}v_1+\dots +\zeta {\rm d}v_{n-1})
$$
which is clearly $\not=0$ when $\zeta \not=0$. Hence $h=0$ defines a closed  analytic submanifold $M$ of dimension $n-1$ of $\Omega=\{(v, \zeta)\ |\ \zeta \not=0\}$. The projection $g : (v, \zeta)\longmapsto v$ when restricted to $M$ has bijective differential at $(0, 1)$. Indeed, this is clear since ${\rm d}g_{(0, 1)}(\partial/\partial \zeta)=n\not=0$. Hence $g$ is an analytic diffeomorphism of an open neighborhood of $(0, 1)$ in $M$ with an open neighborhood of $0$ in $K^{n-1}$. So there exists an integer $a\ge 1$ and an analytic function $f $ on $\{v\ |\ |v_i|\le q^{-a}\}$ such that 
$$
f(v)\equiv 1 (V_p),\qquad h(v, f(v))=0.
$$
This completes the proof of Proposition 3.
\bigskip\noindent
{\bf 4. Proof of Theorem 1.\/} Recall that $U=\{u\in K\ |\ |u|=1\}$.
\bigskip\noindent
{\bf Lemma 1.\/} {\it Let $M$ be a compact open subgroup of $U$. Then there exists an integer $\nu=\nu(M)\ge 1$ with the following property. If $\xi$ is a non-trivial additive character of $K$, then
$$
\int _M\xi(u)\,{\rm d}u=0\qquad ({\rm ord } (\xi)\le -\nu).
$$}
\bigskip\noindent
{\bf Proof.\/} We find an integer $\mu\ge 1$ such that $M_\mu:=1+P^\mu\subset M$. Let $(y_i)$ be a set of representatives for $M/M_\mu$. Then
$$
\int_M\xi(x)\,{\rm d}x=\sum _i \int_{M_\mu}\xi(y_ix)\,{\rm d}x=\sum_i\xi(y_i)\int_{P^\mu}\xi(y_it)\,{\rm d}t.
$$
If $\eta_i(t)=\xi(y_it)$, ${\rm ord } (\eta_i)={\rm ord } (\xi)$ for all $i$, and so
$$
\int_{P^\mu}\eta_i(t)\,{\rm d}t=0\qquad ({\rm ord } (\xi))<-\mu).
$$
Hence, with $\nu=\mu +1$,
$$
\int_M\xi(x)\,{\rm d}x=0\qquad ({\rm ord } (\xi)\le -\mu -1).
$$
\medskip
We shall prove the following. In view of applications to the case of several variables, we prove it under a very general set of conditions. 
\bigskip\noindent
{\bf Proposition 2.\/} {\it Let $K$ be either of characteristic of  $0$ or characteristic $p>0$ but with $p$ not dividing $n$. Let 
$$
h(y)=c_0y^n+c_1y^{n-1}+c_2y^{n-2}+\dots +c_{n-1}y+c_n\qquad (c_j, y\in K)
$$
where $c_0\not=0$ is a fixed constant and the $c_j$ are allowed to vary. Let
$$
I_r=\int _{|y|=q^r}\psi(h(y))\,{\rm d}y.
$$
Then there exist integers $A, r_0\ge 1$ such that 
$$
I_r=0\qquad \left (r\ge r_0, \ |c_i|\le q^{(ir-A)} \ \forall i \right ).
$$}
\bigskip\noindent
{\bf Proof.\/} Replacing $\psi$ by $t\mapsto \psi(c_0t)$ and $c_i$ by $c_ic_0^{-1}$ we may assume that $c_0=1$. Also as $\psi (c_n)$ is constant we may ignore it and so assume $c_n=0$. Let $y=\varpi^{-r}z$ so that
$$
I_r=q^r\int _U\psi(\varpi^{-nr}(F(z : u)))\,{\rm d}z\qquad F(z : u)=z^n+u_1z^{n-1}+\dots +u_{n-1}z
$$
where $u_i=c_i\varpi^{ir}$. With $A$ to be specified later, let $|c_i|\le q^{(ir-A)}$ so that $|u_i|\le q^{-A}\,(1\le i\le n-1)$.
\medskip
Suppose first that $K$ has arbitrary characteristic and that $p$ does not divide $n$. We use Proposition 3.1. Then $|u_i|\le q^{-1}$ for all $i$ if $A=1$. If $A=1$ we make the change of variables $z\longmapsto s=t(z)$ where $t(z)^n=F(z : u)$. Since $nt^{n-1}({\rm d}t/{\rm d}z)=(nz^{n-1}+(n-1)u_1z^{n-2}+\dots +u_{n-1})$ we have $|{\rm d}t/{\rm d}z|=1$ so that
$$
I_r=q^r\int _U\psi(\varpi^{-nr}s^n)\,{\rm d}s.
$$ 
Let $\ell$ be the number of $n^{\rm th}$ roots of unity in $K$ and ${\rm d}^\times s$ the {\it multiplicative\/} Haar measure on $K^\times$. Then ${\rm d}s={\rm d}^\times s$ on $U$ while $s\longmapsto s^n$ is an $\ell$-fold cover of $U$ over $U^n$. Hence
$$
I_r=\ell q^r\int _{U^n}\psi (\varpi^{-nr}\sigma )\,{\rm d}\sigma=\ell q^r\int_{U^n}\xi_r(\sigma)\,{\rm d}\sigma
$$
where $\xi_r$ is the additive character of $K$ defined by $\xi_r(\sigma)=\psi(\varpi^{-nr}\sigma)$.  By Lemma 1, there is an integer $\mu\ge 1$ such that $\int_{U^n}\xi \,{\rm d}\sigma=0$ for all non-trivial additive characters $\xi$ with ${\rm ord } (\xi)\le -\mu$. Now  ${\rm ord } (\xi_r)={\rm ord } (\psi)-nr$ and so 
$$
\int_{U^n}\psi (\varpi^{-nr}\sigma )\,{\rm d}\sigma=0\qquad (nr\ge {\rm ord } (\psi)+\mu).
$$
So, if $r_0:=|{\rm ord } (\psi)|+\mu$ we have $I_r=0$ for $r\ge r_0$. 
\medskip
We must now consider the case when $p$ divides $n$ but $K$ has characteristic $0$. The proof is essentially the same but it now relies on Proposition 3.3. If $a$ is as in that proposition then $|u_i|\le q^{-a}$ for $A=a$. Let $A=a$ and let $W=\{w\}$ be a set of representatives of $U/V_p$. Then 
$$
I_r=q^r\sum_{w\in W}\int_{V_p}\psi(\varpi^{-nr}w^nF_w(z : u))\,{\rm d}z
$$
where
$$
F_w(z : u)=z^n+u_1^\prime \varpi^rz^{n-1}+\dots +u_{n-1}^\prime \varpi^{(n-1)r}z
$$
with
$$
u_i^\prime =c_iw^{-i}\varpi^{ir}, \qquad |u_i^\prime |\le q^{-a} \quad (1\le i\le n-1).
$$
By (b) of that proposition and the choice of $a$ we see that $F_{w,z} : z\longmapsto F_w(z : u)$ is an analytic diffeomorphism $V_p\simeq V_p^n$ with $|{\rm d}F_{w,u}/{\rm d}z|=|n|$. Hence
$$
\int _{V_p}\psi(\varpi^{-nr}w^nF_w(z : u))\,{\rm d}z=|n|^{-1}\int_{V_p^n}\psi(\varpi^{-nr}w^ns)\,{\rm d}s.
$$
We now use Lemma 1. Let $\nu$ be the integer of that lemma when $M=V_p^n$. Since the additive character $\sigma\longmapsto \psi(\varpi^{-nr}w^n\sigma)$ has order equal to ${\rm ord }(\psi)-nr$  we may conclude that the above integral is $0$ if $nr\ge {\rm ord }(\psi)+\nu$. Hence if $r\ge r_0:=|{\rm ord }(\psi)|+\nu$, we have
$$
\int_{V_p}\psi(\varpi^{-nr}w^nF_w(z : u))\,{\rm d}z=0
$$
for all $w$ and hence $I_r=0$ for $r\ge r_0$. This finishes the proof of the entire proposition.  
\medskip\noindent
{\bf Proposition 3.\/} {\it Let $K$ and $h$ be as in Theorem 1. Let
$$
I_r(x)=\int _{|y|=q^r}\psi \big (h(y)-(x,y)\big )\,{\rm d}y.
$$
Then there are integers $B, s_0\ge 1$ such that 
$$
I_r(x)=0 \qquad (r\ge s_0, |x|\le q^{(n-1)r-B}).
$$
\/}
\medskip\noindent
{\bf Proof.\/} We have $(x,y)=\sum _\mu b_\mu(x)y_\mu$ where the $b_\mu$ are linear functions of $x$. We write the set $\{y\in K^m\ \big |\ |y|=q^r\}$ as the disjoint union of sets
$$
E_\mu=\{y\in K^m\ \big |\  |y_\mu|=q^r, |y_i|<q^r(i<\mu) |y_i|\le q^r(i>\mu)\}.
$$
 It is sufficient to prove that for each $\mu$ there are integers $B_\mu, s_\mu\ge 1$ such that
 $$
 \int _{E_\mu} \psi\big (h(y)-(x,y)\big )\,{\rm d}y =0\qquad (r\ge s_\mu, |x|\le q^{(n-1)r-B_\mu}).
 $$
 The integral over $E_\mu$ can be evaluated by first integrating with respect to $y_\mu$ and so it is enough to prove that there are $B_\mu, s_\mu$ such that
 $$
 I_{r, \mu}(x):=\int _{|y_\mu|=q^r}\psi (h(y)-b_\mu(x)y_\mu)\,{\rm d}y_\mu=0\qquad (r\ge s_\mu, |x|\le q^{(n-1)r-B_\mu})
 $$
for fixed $y_j(j\not=\mu)$ with $|y_j|<q^r(j<\mu), |y_j|\le q^r (j>\mu)$. 
\medskip
Now
$$
h(y)=a_\mu y_\mu^n+c_1y_\mu^{n-1}+\dots +c_{n-2}y_\mu^{2}+(c_{n-1}-b_\mu(x))y_\mu+c_n
$$
where $c_i$ is a polynomial in the $y_j(j\not=\mu)$ of degree $\le i-1$. Hence there is an integer $C\ge 1$ such that for $1\le i\le n-1$,
$$
|c_i|\le q^{(i-1)r+C}\qquad (|y|\le q^r).
$$
On the other hand $|b_\mu(x)|\le q^D|x|$ for some integer $D\ge 1$ for all $x$ so that 
$$
|c_{n-1}-b_\mu(x)|\le \max (q^{(n-2)r+C}, q^D|x|).
$$
Let $A, r_0$ be as in Proposition 2. Then that proposition implies that $I_{r, \mu}(x)=0$ if 
$$
r\ge r_0,\quad (i-1)r+C\le ir-A \quad (1\le i\le n-2)
$$
and 
$$
\max (q^{(n-2)r+C}, q^D|x|)\le q^{(n-1)r-A}.
$$
If we put $B_\mu=D+A$ and $s_\mu=\max (r_0, B_\mu+C)$ we satisfy the above requirements when $r\ge s_\mu$ and $|x|\le q^{(n-1)r-B_\mu}$, and so we are done.
\bigskip\noindent
{\bf Completion of the proof of Theorem 1.\/} We have actually proved $(\ast\ast)$ of \S1 in the stronger form discussed there and so all statements of Theorem 1 are proved.
\medskip
For the converse, let $m=1$ and let $h$ be linear, say $h(y)=by$. Then 
$$m
A_h(f)=\int_K\psi(h(y))\widehat f(y)\,{\rm d}y=\int _K\psi(by)\widehat f(y)\,{\rm d}y=f(b)
$$
so that $A_h$ is the delta function at $b$.
\bigskip\noindent
{\bf Remark.\/} It is interesting that the bound on $A_h(x)$ is the same as in the real case [FV].
\bigskip\noindent
{\bf 5. Completion of the proof of Theorem 2 in characteristic $p$.\/} We now consider the case $K=K_f={\f}_q[[T]][T^{-1}]$ and $p$ divides $n$. Here we assume that $m=1$ and use the notation developed in \S 2. 
\medskip
We first note that for $a,b\in {\f}_q$, 
$$
{\rm Tr}(ab^p)={\rm Tr}(a^{p^{-1}}b).
$$
This follows from the fact that ${\rm Tr}(c)={\rm Tr}(c^p)$ for all $c\in {\f}_q$. Indeed, observe that the Galois group of ${\f}_q$ over ${\f}_p$ is cyclic of order $f$ generated by the Frobenius map $x\longmapsto x^p$, and so, as  ${\rm Tr}(c)$ is the sum of the elements of the Galois orbit of $c$ and as $c^p$ is in the same orbit, the traces of $c$ and $c^p$ are the same.
\medskip
Recall that for $c\in K_f$,
$$
 Qc=c^\sharp =\sum_jc^\sharp (j)T^j,\qquad c^\sharp(j)=(c(-1+p(j+1)))^{p^{-1}}\qquad (c\in K_f).
$$
\bigskip\noindent
{\bf Lemma 1.\/} {\it Let $c, u\in K_f$, $r\ge 1, (m,p)=1$. Then 
$$
{\rm Tr}({\rm Res}(cu^{mp^r}))={\rm Tr}({\rm Res}((Q^rc)u^m))\qquad
\psi(cu^{mp^r})=\psi ((Q^rc)u^m).
$$}
\bigskip\noindent
{\bf Proof.\/} We have
$$
\eqalign {{\rm Tr}\bigg ({\rm Res}(cu^p)\bigg )&={\rm Tr}\bigg ({\sum}_{r+\ell p=-1}c_ru^p_\ell\bigg )={\rm Tr}\bigg ({\sum}_{r+\ell p=-1}c_r^{p^{-1}}u_\ell\bigg)\cr
&={\rm Tr}\bigg (\sum _\ell c_{-1-\ell p}^{p^{-1}}u_\ell \bigg )={\rm Tr}\bigg (\sum _\ell c^{\sharp}(-\ell -1)u_\ell\bigg )\cr
&={\rm Tr}\bigg ({\rm Res}((Qc)u)\bigg ).\cr }
$$
Repeated application of this gives the result.
\medskip
It follows from the lemma that $\psi(h)=\psi(h^\sharp)$. This finishes the proof of Theorem 2 except for the converse statement. If $h^\sharp$ is of degree $\le 1$, say $h^\sharp =by$, then for $f$ a Schwartz-Bruhat function on $K$, 
$$
A_h(f)=\int_K\psi(h(y))\widehat f(y)\,{\rm d}y=\int _K\psi(by)\widehat f(y)\,{\rm d}y=f(b)
$$
so that $A_h$ is the delta function at $b$.
\medskip
As an example, let $h(y)=cy^3$ on $K={\f}_3[[T]][T^{-1}]$ where $c\in K$ and non-zero. Then $h^\sharp(y)=c^{\sharp}y$ and so $A_h$ is the delta function at $c^\sharp$. But if we take $h(y)=cy^3+ay^2$ where $a\not=0$, then $h^\sharp (y)=ay^2+c^\sharp y$ and so $A_h$ {\it is\/} a locally constant function which is $O(|x|^{1/2})$ at infinity on $K$.
\bigskip\noindent
{\bf 6. Airy integrals associated to compact $p$-adic Lie groups.\/} If $G$ is a compact $p$-adic Lie group with Lie algebra $\frak g$, it is natural to ask if some of the $G$-invariant polynomials on $\frak g$ have the Airy property. The example below suggests that this may be the case.
\medskip
This example is the analogue of ${\rm SO}(3)$ over the $p$-adic field. We work over a local field $K$ of characteristic $\not=2$ and use the notation of \S 1. Let $\alpha\in R$ be such that $\alpha$ is not a square mod $P$. Then $\alpha$ is not a square in $K$ either by Hensel's lemma. We consider the algebra ${\hh}$ with generators ${\bf i}, {\bf j}$ and relations
$$
{\bf i}^2=\alpha,\quad {\bf j}^2=\varpi,\quad {\bf i}{\bf j}=-{\bf j}{\bf i}={\bf k}\,(\hbox {say}).
$$
It is known that ${\hh}$ is a division algebra with center $K$ and that it is, up to isomorphism, the only $4$-dimensional division algebra over $K$ with center $K$. On ${\hh}$ we have the trace ${\rm Tr}$ and norm $N$ given by
$$ 
{\rm Tr}(x)=2a_0,\quad N(x)=x\bar x=a_0^2-\alpha a_1^2-\varpi (a_2^2-\alpha a_3^2)\quad (x=a_0+a_1{\bf i}+a_2{\bf j}+a_3{\bf k})
$$ 
where $x\mapsto \bar x$ is the involutive anti-automorphism of ${\hh}$ such that 
$$
\bar x=a_0-a_1{\bf i}-a_2{\bf j}-a_3{\bf k}\quad \hbox {for } x=a_0+a_1{\bf i}+a_2{\bf j}+a_3{\bf k}.
$$
We write $|x|=|N(x)|^{1/2}_{{\q}_p}$ which is a non-archimedean norm over ${\hh}$. It is known [W] that $N$ maps ${\hh}\setminus \{0\}$ onto $K^\times$. Hence the norm values on ${\hh}$ consist of $0$ and the numbers $q^r (r\in (1/2){\z})$.
\medskip
Let $L=K+K{\bf i}$. Then $L$ is a subfield of ${\hh}$ and is the quadratic extension $K(\sqrt {\alpha})$. The algebra ${\hh}$ becomes isomorphic over $L$ to the full matrix algebra in dimension $2$. To see this write ${\hh}=L\oplus {\bf j}L$. We consider the action $ \lambda $ of ${\hh}$ on itself by left multiplication. In the basis $\{1, {\bf j}\}$ we have
$$
\lambda (u+{\bf j}v)=\pmatrix {u&\varpi \bar v\cr v&\bar u\cr}\quad 
\lambda : 1\mapsto I,\quad {\bf i}\mapsto \pmatrix {{\bf i}&0\cr 0&-{\bf i}\cr}\quad {\bf j}\mapsto \pmatrix {0&\varpi \cr 1&0.\cr}. 
$$
Thus $\lambda$ extends to an isomorphism of $L\otimes _{K}{\hh}$ with the full matrix algebra over $L$. We have
$$
{\rm Tr}(x)={\rm Tr}(\lambda (x))\qquad N(x)=\det (\lambda (x)).
$$
\medskip
The fact that ${\hh}$ is a division algebra means that $x=0\Leftrightarrow x\bar x=0\Leftrightarrow N(x)=0$. Hence $N$ is a quadratic form on ${\hh}$ which is non-zero everywhere on ${\hh}\setminus \{0\}$, i.e., it is {\it anisotropic\/}. The bilinear form corresponding to $N$ is $B(x, y)={\rm Tr}(x\bar y)$. Let $\frak g$ be the subspace of ${\hh}$ of elements of trace zero, i.e.,
$$
\frak g=\{x\in {\hh}\ |\ {\rm Tr}(x)=0\}.
$$
Then $\frak g$ is a Lie algebra and $B(x,y)=-{\rm Tr}(xy)$ for $x,y\in \frak g$ since $\bar u=-u$ for $u\in \frak g$. The special orthogonal group ${\bf {\rm SO}}(B)$ is algebraic and defined over $K$. We write $G$ for its group of points over $K$. Since $N$ is anisotropic it follows that $G$ is compact. Clearly $\frak g$ is its Lie algebra. We define the polynomials $p_r$ on $\frak g$ by
$$
p_r(x)={\rm Tr}(x^r)\quad (x\in \frak g).
$$
 If $0\not=x\in \frak g$ the eigenvalues of $\lambda(x)$ cannot be $0$ as  $\det (\lambda(x))=N(x)\not=0$. Hence $\lambda(x)$ is semisimple and we may write its eigenvalues as $\pm \varepsilon (x)$. Then $p_r(x)=0$ for $r$ odd while $p_{2s}(x)=2\varepsilon (x)^{2s}=2(-1)^s N(x)^s$. Thus
 $$
 p_r(x)=\cases {0 & if $r$ is odd\cr 2(-1)^sN(x)^s & if $r=2s$.\cr}
 $$
Thus the $p_{2s}$ are invariant under $G$. Note that $p_{2s}(u)\not=0$ for $u\not=0$.
\bigskip\noindent
{\bf Theorem 1.\/} {\it The polynomials $p_{2s}$ have the Airy property. The associated Airy function $A_{p_{2s}}$ is $O\big (|x|^{3/(2s-1)}\big)$ at infinity on $\frak g$.\/}
\bigskip\noindent
{\bf Remark.\/} The method of proof is similar to that for the case of the real field but there are new features that make it difficult at this time to extend this result to other compact $p$-adic Lie groups. The growth estimate at infinity on $\frak g$ is surprisingly the same as in the real case.
\medskip
The proof depends on the following two lemmas, the first of which is just the Weyl integration formula. For any $z\not=0$ in $\frak g$, let $\frak g(z)$ be the set of all elements of $\frak g$ conjugate under $G$ to a non-zero multiple of $z$. Now two elements $y, y'$ of $\frak g$ are conjugate under $G$ if and only if $N(y)=N(y')$ and so $\frak g(z)$ consists of all elements $y$ such that $N(y)\in N(z){K^\times }^2$. Hence $\frak g(z)\subset \frak g\setminus \{0\}$ and is open, invariant, and closed in $\frak g\setminus \{0\}$. Let $G_z$ be the stabilizer of $z$ in $G$ and $\overline G=G/G_z$. We write ${\rm d}g$ (resp.$\,{\rm d}\bar g$) for the normalized Haar measure on $G$ (resp.$\,\overline G$).
\bigskip\noindent
{\bf Lemma 1.\/} {\it There is a constant $\gamma=\gamma (z)>0$ with the following property. For any $f\in {\ss}{\bb}(\frak g)$,
$$
\int _{\frak g(z)}f(y)\,{\rm d}y=\gamma\int _{K^\times} |t|^2\varphi_f(tz)\,{\rm d}t\qquad \varphi_f (tz)=\int _{G}f(tg{\cdot}z)\,{\rm d}g.
$$}
\bigskip\noindent
{\bf Proof.\/} The proof is similar to the real case. We set up the analytic map $\varphi$ of $K^{\times}\times \overline G\longrightarrow \frak g(z)$ given by $\varphi : (t, \bar g)\longmapsto tg{\cdot}z$. Clearly $\varphi$ is surjective. We claim that $\varphi$ is $2:1$ and $d\varphi$ is bijective. If $tg{\cdot}z=t'g'{\cdot}z$, then, taking norms, $t^2=t'^2$, and $t=t'$ gives $\bar g=\bar {g'}$. On the other hand, if $h\in {\rm SO}(\frak g)$ is such that $h{\cdot}z=-z$ (which is possible since $N(z)=N(-z)$), then $hG_z$ is uniquely determined. Moreover $h$ normalizes $G_z$ and so the map $\bar g\mapsto w(\bar g):=\overline {gh}$ is well defined and involutive. So, when $t'=-t$ we have $g'=w(\bar g)$. The fibre of $\varphi$ is $\{(t,\bar g), (-t, w(\bar g)\}$.  
\medskip 
For proving the bijectivity of ${\rm d}\varphi$, note that $\varphi$ commutes with the actions of $G$ on $\overline G$ and $\frak g(z)$, so that it is enough to check it at the points $(t, \bar 1)$. We shall identify the Lie algebra of $G$ with $\frak g$ acting on itself by the adjoint action. Let $\{z_1=z, z_2, z_3\}$ be an orthogonal basis for $\frak g$; then the commutation rules are easily checked to be $[z_i, z_j]=e_{ijk}z_k$ where $(ijk)$ is an even permutation of $(123)$ and  $e_{ijk}\not=0$. The tangent space to $\overline G$ at $\bar 1$ is identified with $\frak g/K{\cdot}z\simeq Kz_2\oplus Kz_3$ and so the tangent space of ${\q}_p^\times \times \overline G$ at $(t,1)$ is identified with $\frak g$ by sending ${\rm d}/{\rm d}\tau$ to $z$. If $D_t=({\rm d}\varphi)_{(t,\bar 1)}$ then 
$$
D_t : z_1\mapsto ({\rm d}/{\rm d}\tau )_{\tau=0}((t+\tau)z_1)=z_1,\  z_2\mapsto t[z_2, z_1]=-e_{123}tz_3,\  z_3\mapsto e_{312}tz_2
$$
so that $\det (D_t)=ct^2$ where $c\not=0$. This proves the bijectivity and at the same time shows that the Jacobian of the map is $ct^2$. The formula of Lemma 1 is now clear.  
\medskip
We use the form $(x, y):={\rm Tr}(xy)$ to define the Fourier transform on $\frak g$. Note that $|(x, y)|\le \alpha |x||y|$ for all $x,y\in \frak g$ where $\alpha >0$ is a constant.
\bigskip\noindent
{\bf Lemma 2.\/} {\it We can find integers $r_0=r_0(z), A_0=A_0(z)\ge 1$ such that
$$
I_r(z, x):=\int _{\{|y|=q^r\}\cap \frak g(z)}\psi \big (p_{2s}(y)-(x,y)\big )\,{\rm d}y=0
$$
for all $r\ge r_0, x\in \frak g$ with $|x|\le q^{(2s-1)r-A_0}$.\/}
\bigskip\noindent
{\bf Proof.\/} For $r$ any half integer $>1$ we apply Lemma 1 to the function
$$
f(y)=\chi(|y|=q^r) \psi \big (p_{2s}(y)-(x,y)\big ).
$$
Here $\chi$ is the characteristic function of the set indicated. If $|z|=q^{a(z)}$ then $|tg{\cdot}z|=q^r\Leftrightarrow |t|=q^{r-a(z)}$ (note that this is possible only if $r-a(z)$ is an integer). Hence
$$
\int _{\{|y|=q^r\}\cap \frak g(z)}\psi \big (p_{2s}(y)-(x,y)\big )\,{\rm d}y
$$
is equal to 
$$
\int _{|t|=q^{r-a(z)}}|t|^2\bigg (\int _G \psi \big (p_{2s}(tz)-(tx, g{\cdot}z)\big )\,{\rm d}g\bigg )\,{\rm d}t.
$$
Now $p_{2s}(tz)=\beta t^{2s}$ where $\beta=p_{2s}(z)\not=0$ while $t(x, g{\cdot}z)=:\gamma (x,g)t$ where $|\gamma (x, g)|\le \alpha |x||z|$. Hence the above integral, {\it after inverting the order of integration\/}, can be written as 
$$
q^{2(r-a(z))}\int _G \bigg (\int _{|t|=q^{r-a(z)}}\psi \big (\beta t^{2s}-\gamma (x,g)t\big )\,{\rm d}t\bigg )\,{\rm d}g.
$$
By Proposition 3 of \S 4 (with $m=1$) with $s_0, B$ as in that proposition, the inner integral above is $0$ if $r-a(z)\ge s_0, |\gamma (x,g)|\le q^{(2s-1)(r-a(z))-B}$. As $|\gamma(x, g)|\le \alpha |x||z|$, this condition is satisfied if $|x|\le \alpha^{-1}|z|^{-1}q^{(2s-1)(r-a(z))-B}$ which is {\it independent of $g\in G$\/}. If we now choose $r_0$ such that $r_0\ge s_0+a(z)$ and $A_0$ such that $q^{-A_0}\le \alpha ^{-1}|z|^{-1}q^{-(2s-1)a(z)-B}$ we see that the inner integral is $0$ if $r\ge r_0, |x|\le q^{(2s-1)r-A_0}$ which is again a condition independent of $g\in G$. The lemma is now clear.
\bigskip\noindent
{\bf Proof of Theorem 1.\/} We now observe that there is a {\it finite\/} set $F\subset   \frak g$ such that $\frak g\setminus \{0\}$ is the {\it disjoint\/} union of the $\frak g(z)$ for $z\in F$. To see this note that $N(tu)=t^2N(u)$ for $t\in K^\times, u\in \frak g$ and so the image of the norm function on the non-zero part of $\frak g$ is a union of cosets $K^\times/{K^\times}^2$. But $K^\times/{K^\times}^2$ is finite and so there is a finite set $F\subset \frak g$ such that for any $y\in \frak g$ there are $z\in F, t\in K^\times$ such that $N(tz)=N(y)$, i.e., $y\in \frak g(z)$. Hence
$$
I_r(x):=\int _{\{|y|=q^r\}}\psi \big (p_{2s}(y)-(x, y)\big )\,{\rm d}y=
\sum _{z\in F}I_r(z, x).
$$  
By Lemma 2 we can then find positive integers $r_1, A_1$ such that $I_r(x)=0$ for $r\ge r_1, |x|\le q^{r-A_1}$. From this point onwards the argument is the same as before and leads to Theorem 1.
\bigskip\noindent
{\bf Acknowledgments.\/}  R.$\,$N.$\,$F. would like to thank Alexander Bobenko for inviting him to visit  the Technische Universit\"at, Berlin;  this visit was supported by an ENIGMA fellowship.  He would also like to thank the Max-Planck-Institut f\"ur Mathematik, Bonn, for their hospitality during his visit in 2008.  V.$\,$S.$\,$V. is grateful to the INFN, Genova, Italy for their hospitality in September 2008 for a stay during which a part of the results of this paper were worked out. D.$\,$W. would like to thank the Department of Mathematics at UCLA for their hospitality. We also like to thank Professor Pierre Deligne for his comments.
\vskip 1 true in
\centerline { \bf References}
\vskip 0.5 true in
\item {[A]} G. B. Airy, {\it On the intensity of light in the neighbourhood of a caustic\/}, Trans. Camb. Phil. Soc. 6 (1838), pp. 379--403. 
\medskip \item {[FV]} Rahul N. Fernandez, and V. S. Varadarajan {\it Matrix Airy functions for compact Lie Groups\/}, To appear in Int. Jour. Math.
\medskip\item {[H]} Harish-Chandra, {\it Harmonic analysis on reductive $p$-adic groups\/}, Proceedings of Symposia in Pure Mathematics, Vol. XXVI, Amer. Math. Soc. Providence, R. I., 1973, pp. 167--192. See also {\it Harish-Chandra Collected Papers\/}, Vol. IV, pp. 75--100.
\medskip\item {[K]} Maxim Kontsevich, {\it Intersection theory on the moduli space of curves and the matrix Airy function\/}, Comm. Math. Phys. 147 (1992), no.1, pp. 1--23. 
\medskip\item {[L]} Edouard Looijenga, {\it Intersection theory on Deligne-Mumford compactifications\/}, S\'eminaire BOURBAKI, n$^\circ$768, (1992-93). 
\medskip\item {[V1]} I. V. Volovich, {\it Number theory as the ultimate theory\/}, CERN preprint CERN-TH.4791/87, 1987.
\medskip\item {[V2]} I. V. Volovich, {\it $p$-adic string\/}, Class. Quantum Grav. 4(1987) L83--L87.
\medskip\item {[VVZ]} V. S. Vladimirov, I. V. Volovich, and E. I. Zelenov, {\it $p$--adic Analysis and Mathematical Physics,\/} World Scientific, 1994. 
\medskip\item {[Va]} V. S. Varadarajan, {\it Multipliers for the symmetry groups of $p$-adic spacetime\/}, To appear in $p$-Adic Numbers, Ultrametric Analysis, and Application. 
\medskip\item {[Vi]} J. Virtanen, {\it Structure of Elementary Particles in Non-Archimedean Spacetime\/}, Thesis, UCLA, 2009.
\medskip\item {[W]} Andr\'e Weil, {\it Basic Number Theory\/}, Reprint of the second (1973) edition. Classics in Mathematics. Springer-Verlag, Berlin, 1995.
\medskip\item {[Wi]} Edward Witten, {\it Two-dimensional gravity and intersection theory on moduli space\/}, Surveys in differential geometry (Cambridge, MA, 1990), 243--310, Lehigh Univ., Bethlehem, PA, 1991. 
\vskip 0.5 true in\noindent
{\mysmall Rahul~N.~Fernandez, Institut f\"ur Mathematik MA 3--2, Technische Universit\"at Berlin, Strasse des 17.~Juni 136,
10623 Berlin, Germany, {\eightit fernand@math.tu-berlin.de.} \smallskip\noindent
V. S. Varadarajan, Department of Mathematics, UCLA, Los Angeles, CA 90095-1555, USA, {\eightit vsv@math.ucla.edu}
\smallskip\noindent
David Weisbart, Department of Mathematics, UCLA, Los Angeles, CA 90095-1555, USA, {\eightit dweisbar@math.ucla.edu}}
\medskip

\bye